\documentclass[twocolumn,twocolappendix]{aastex63}

\received{2021 February 19}
\revised{2021 June 25}
\accepted{2021 June 27}
\published{2021 September 23}
\shorttitle{GAL-CLUS-022058s}
\shortauthors{D\'\i az S\'anchez et al.}

\begin{document}

\title{The Einstein ring GAL-CLUS-022058s: a Lensed Ultrabright Submillimeter Galaxy at z=1.4796}

\correspondingauthor{A. D\'{\i}az-S\'anchez
  }
\email{Anastasio.Diaz@upct.es}

\author[0000-0003-0748-4768]{A. D\'{\i}az-S\'anchez
}

\affil{
Departamento F\'\i sica Aplicada, Universidad Polit\'ecnica de Cartagena, Campus Muralla del Mar, 30202 Cartagena, Murcia, Spain\\
}

\author[0000-0001-7147-3575]{H. Dannerbauer}
\affiliation{
Instituto de Astrof\'\i sica de Canarias,  V\'\i a L\'actea, La Laguna 38200, Spain\\
}
\affiliation{
Departamento de Astrof\'\i sica de la Universidad de La Laguna, Avda. 
Francisco S\'anchez, La Laguna, 38200, Spain
}

\author[0000-0002-3187-1648]{N. Sulzenauer}
\affiliation{
Max-Planck-Institut für Radioastronomie, Auf dem Hügel 69, 53121 Bonn, Germany\\
}

\author[0000-0003-0618-0615]{S. Iglesias-Groth}
\affiliation{
Instituto de Astrof\'\i sica de Canarias,  V\'\i a L\'actea, La Laguna 38200, Spain\\
}
\affiliation{
Departamento de Astrof\'\i sica de la Universidad de La Laguna, Avda. 
Francisco S\'anchez, La Laguna, 38200, Spain
}

\author[0000-0003-3767-7085]{R. Rebolo}
\affiliation{
Instituto de Astrof\'\i sica de Canarias,  V\'\i a L\'actea, La Laguna 38200, Spain\\
}
\affiliation{
Departamento de Astrof\'\i sica de la Universidad de La Laguna, Avda. 
Francisco S\'anchez, La Laguna, 38200, Spain
}
\affiliation{
Consejo Superior de Investigaciones Cient\'\i cas, Spain \\
}



\begin{abstract}

We report an ultra-bright lensed submillimeter galaxy at $z_{spec}=1.4796$, identified as a result of a full-sky cross-correlation of the {\it AllWISE} and {\it Planck} compact source catalogs aimed to search for bright submillimeter galaxies at $z \sim 1.5-2.8$.
APEX/LABOCA observations of the candidate galaxy reveal a source with flux, S$_{870 \mu m}= 54\pm 8$ mJy. The position of the APEX source coincides with the position of the {\it AllWISE} mid-IR source, and with the Einstein ring GAL-CLUS-022058s, observed with the {\it HST}. Archival VLT/FORS observations reveal the redshift of this Einstein ring, $z_{spec}=1.4796$, and detection of the CO(5-4) line at $z_{spec} = 1.4802$ with APEX/nFLASH230 confirms the redshift of the submillimeter emission. The lensed source appears to be gravitationally magnified by a massive foreground galaxy cluster lens at $z = 0.36$. We use Lenstool to model the gravitational lensing, which is near to a ``fold arc'' configuration for an elliptical mass distribution of the central halo, where four images of the lensed galaxy are seen; the mean magnification is $\mu_{\rm L} =18\pm 4$. We have determined an intrinsic rest-frame infrared luminosity of $L_{\rm IR} \approx 10^{12} L_\sun $ and a likely star formation rate of $\sim 70-170$ $M_\sun\ yr^{-1}$. The molecular gas mass is $M_{\rm mol} \sim 2.6 \times 10^{10} M_\sun$ and the gas fraction is $f = 0.34\pm 0.07$. We also obtain a stellar mass log$(M_\ast/M_\sun) = 10.7 \pm 0.1$ and a specific star formation rate log$(sSFR/Gyr^{-1})=0.15 \pm 0.03$. 
This galaxy lies on the so-called main sequence of star-forming galaxies at this redshift.
\end{abstract}



\section{Introduction}

Submillimeter-selected galaxies \citep[see][for a review]{2014PhR...541...45C} allow the study of the key phases of the formation and evolution of the most massive galaxies in the distant universe. The peak of the submillimeter galaxy (SMG) redshift distribution is at $z\approx 1.7-2.8$ \citep{2005ApJ...622..772C} coinciding with the highest cosmic star-formation rate density. Furthermore, the contribution to the star-formation history of the most luminous, dusty star-forming galaxies steeply increases during earlier time \citep{2005ApJ...632..169L}. Forming stars in compact starbursts with extremely high rates \citep[see e.g.][]{2005MNRAS.359.1165G}, exceeding thousand solar masses per year, could explain the rapid build-up of present-day giant ellipticals \citep[see][]{2013ApJ...772..137I}. Surveys with the {\it Herschel} space observatory, such as H-ATLAS \citep{2010Sci...330..800N}, HerMES \citep{2012ApJ...753...23M}, and with the South Pole Telescope \citep[SPT;][]{2013Natur.495..344V,2013ApJ...767...88W} have provided an increasing number of the most luminous of these dusty star-forming galaxies. But due to their compact sizes, resolved observations of their star-forming Interstelar medium (ISM) remained limited, even with contemporary mm/submm facilities (see e.g. \cite{2015MNRAS.452.2258D}). In addition, high dust attenuation at optical and near-infrared (NIR) bands impede the identification of SMGs \citep{2002ApJ...573..473D} and further analysis of gas lines in SMGs at these wavelengths.

Strong gravitational lensing, especially with galaxy clusters as deflectors, enables efficient detection of SMGs \citep{2010Sci...330..800N}.
The {\it Herschel} and the {\it Planck} space mission have provided detections of some of the brightest lensed SMGs at the whole sky. In \cite{2021ApJ...908..192S} they carried out ALMA observations of lensed SMGs selected by {\it Herschel}, where 29 sources with $z \sim 1.0-3.2$ in 20 fields of massive galaxy clusters have been detected, with a median total IR luminosity $\mu_L L_{\rm IR}= 10^{12.92\pm0.07} L_\odot$. Follow up observations of 24 strongly lensed star-forming galaxies between $z \sim 1.0-3.5$ from the {\it Planck} satellite have been presented in \cite{2021ApJ...908...95H} with $\mu_L L_{\rm IR} \sim 10^{13-14.6} L_\odot $. 
The Cosmic Eyebrow \citep{2017ApJ...843L..22D} is one of the brightest SMGs in the sky, it is a lensed dusty star-forming high redshift galaxy at the peak epoch of the star formation in the Universe, at $z=2.04$. Observations of this galaxy with IRAM/NOEMA at the redshifted CO(3-2) line led to the brightest detection of this line in a SMG confirming its submillimeter nature and extreme molecular gas properties \citep{2019AJ....158...34D}.
The source magnifications permit resolved investigations of giant molecular clouds at a scale of $\sim$100 pc. Together with high-precision lensing analysis, observations of the molecular media of lensed galaxies allow to test models on the process of star-formation in the early universe \citep{2018NatAs...2...76C,2019NatAs...3.1115D}. Evidence of significant star-forming clumps down to a spatial scale of 80 pc has been found in the Cosmic Snake at z=1.04 \citep{2019NatAs...3.1115D} but it is lacking in the Cosmic Eyelash at z= 2.3 \citep{2020MNRAS.495L...1I}.
Although non-lensed SMGs with high emission in the submillimeter are obscured in the NIR and optical bands, lensing magnification also helps to study the rest-frame UV and optical lines \citep[see e.g.][]{2019MNRAS.488.5862A}.

Based on data from the {\it Planck} and {\it WISE} space missions, we report here the discovery of {\it WISE} J022057.56-383311.4, a magnified SMG at $z=1.48$. VLT archival data have allowed us to obtain the redshift, and follow up observations with APEX have confirmed the optical redshift and the strong submillimiter fluxes observed with Planck coming from the molecular gas. The paper is structured as follows:
we first summarize our method for searching lensed SMGs in section \ref{sec:search}.
In section \ref{sec:obser}, we present the observations and data reduction for APEX observations, archival {\it HST} observations, archival VLT spectroscopy, archival GEMINI spectroscopy, archival radio sources, and archival {\it Chandra} observations. We analyze the data in section \ref{sec:ana}, we first study the galaxy cluster properties, then we model the gravitational lensing, after that the spectral energy density (SED) is presented and some properties of the SMG are derived, and finally the CO lines are studied. In section \ref{sec:dis}, we discuss the physical properties of the SMG, and in section \ref{sec:concl} some conclusions are presented. We adopt a flat $\Lambda$CDM cosmology from \cite{2016A&A...594A..13P} with $H_0=68$ km s$^{-1}$ Mpc$^{-1}$, $\Omega_m=0.31$ and $\Omega_\Lambda = 1 - \Omega_m$.

\section{The search} \label{sec:search}

As described in \cite{2017ApJ...843L..22D}, we carried out a cross-matching between the {\it AllWISE}\footnote{\url{http://wise2.ipac.caltech.edu/docs/release/allwise/}} and {\it Planck}\footnote{\url{http://pla.esac.esa.int/pla/\#home}} full-sky compact source catalogs in order to find bright lensed SMGs in the full-sky. 
First we built a full-sky selection of galaxies verifying the color criteria $W1-W2>0.8$, $W2-W3<2.4$, $W3-W4>3.5$, taken from the Cosmic Eyelash referenced SED \citep{2010Natur.464..733S}, and detection at $S/N>5$ in the four bands of the {\it AllWISE} catalog, we limited the sample to galactic latitudes $|b| \geq 20^{o}$. Only SMGs at the peak epoch of the star formation in the Universe ($z=1.5-2.8$) are expected to fulfill these criteria \citep{2017MNRAS.467..330I}. Then, we requested the {\it AllWISE} selected objects to have a {\it Planck} source detected within 5 arcmin with submillimeter flux ratios consistent with those expected for SMGs at $z=1.5-3$. We removed the sources which clearly showed contamination by galactic dust emission.
This led to the identification of two sources {\it WISE} J132934.18+224327.3, nicknamed Cosmic Eyebrow, and {\it WISE} J022057.56-383311.4. This second source is the most likely counterpart of the submillimeter {\it Planck} source PCCS2 857 G249.95-68.09 and a very promising candidate as an ultra-bright SMG.  

A subsequent search in the {\it HST}\footnote{\url{http://archive.stsci.edu/}} archive provided images of a lensed galaxy at the same region where {\it WISE} J022057.56-383311.4 is located. This source is within 30 arcsec from the centroid of the {\it Planck} source and shows a spectacular strongly gravitational lensed galaxy with arc structures larger than 25 arcsec, the contours in the {\it WISE} image are elongated along the main arc structure. A massive galaxy cluster at $z=0.36$ is lensing this galaxy. This galaxy cluster and the lensed SMG have been observed previously with {\it HST}, VLT/FORS2, GEMINI-S and {\it Chandra}.

\section{Observations and data reduction} \label{sec:obser}

\begin{figure*}
\gridline{
          \fig{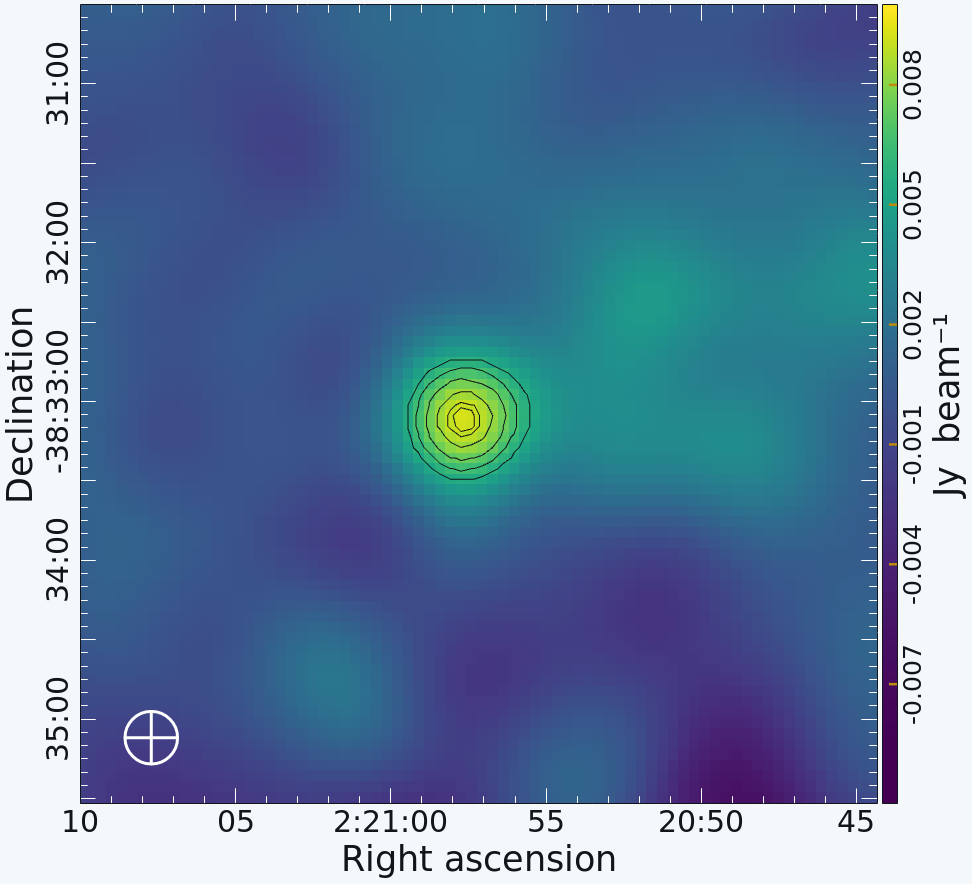}{0.25\textwidth}{(a)}
          \fig{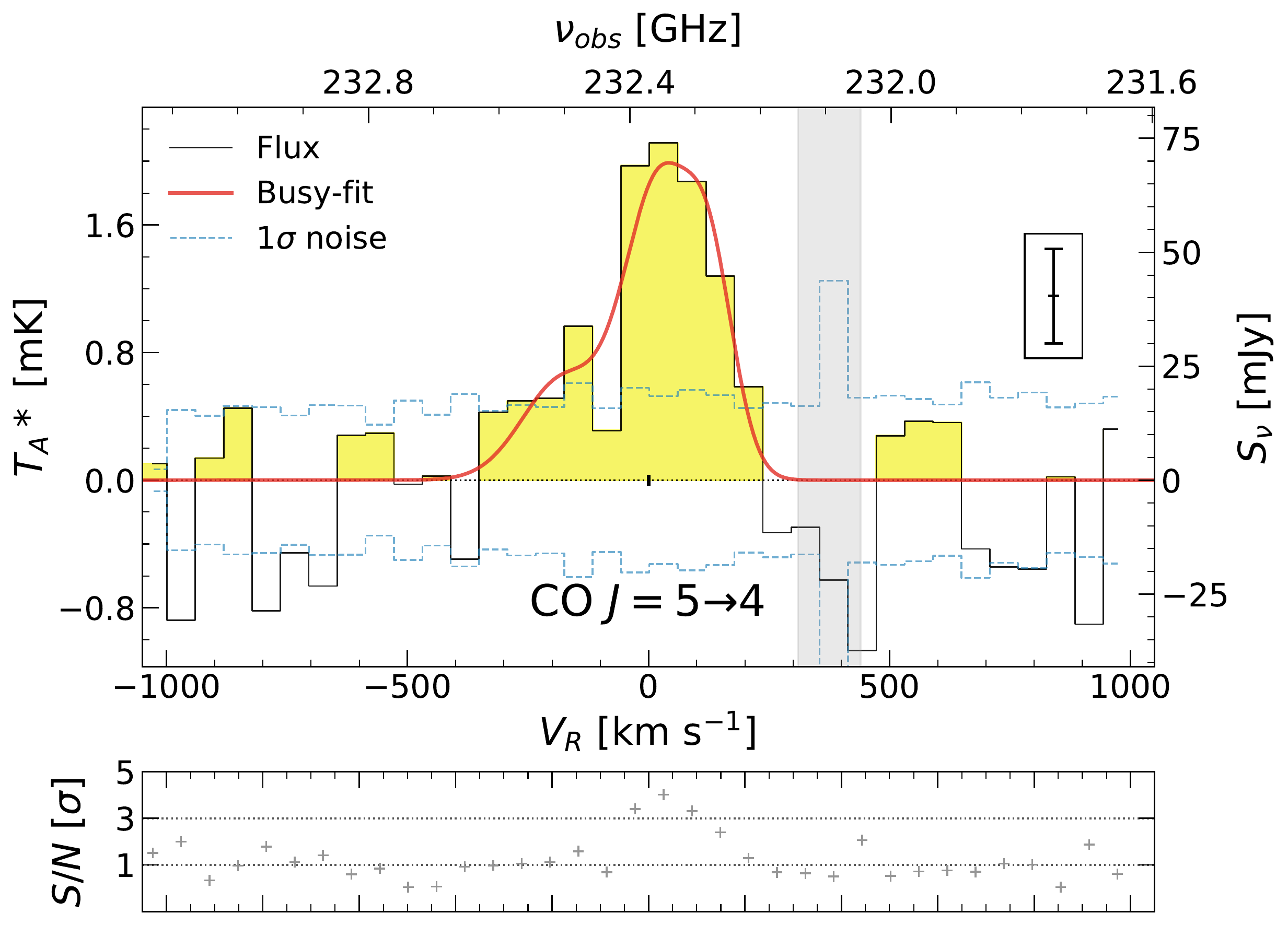}{0.35\textwidth}{(b)}
          \fig{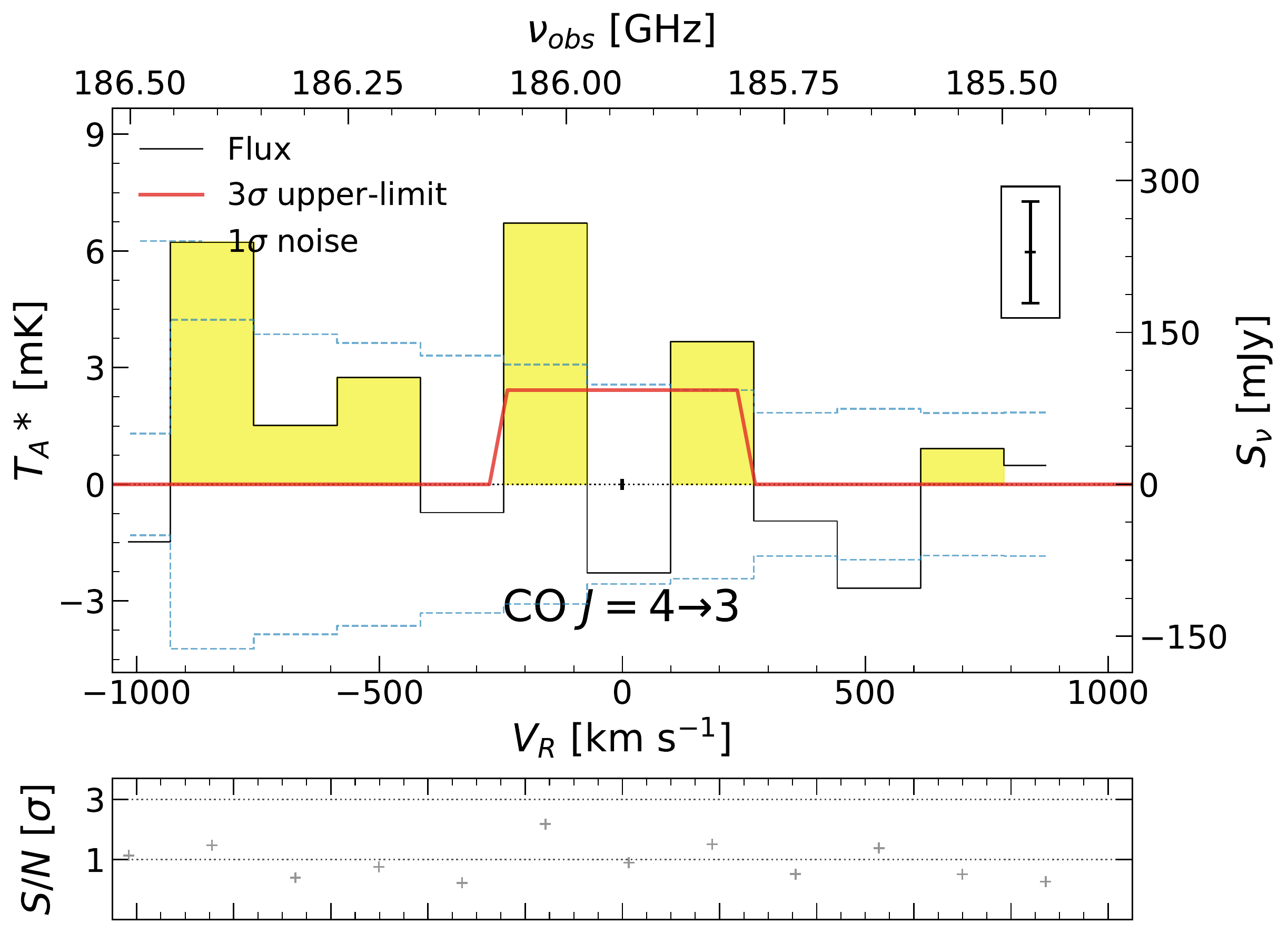}{0.35\textwidth}{(c)}
          }
\caption{(a) APEX/LABOCA signal map at 870 $\mu$m (345 GHz), we show a field of $5\times 5$ arcmin$^2$, north is up; east is left. Contour levels are $3\sigma$, $3.5\sigma$, $4\sigma$, $4.5\sigma$, $4.8\sigma$, and $4.9\sigma$. The beam size is also shown bottom-left in white.
(b) APEX/nFLASH230 spectrum centered at the redshifted CO(5-4) emission line frequencies of the lensed source, obtaining a spectroscopic redshift $z=1.4802$. 
On the vertical axis, two representations of flux density are shown. Equally, the horizontal axis provides information on the channel frequency
and radial velocity in the rest-frame of the source. Additionally, careful noise reduction is performed via the rms selection function and the 
noise content per binned channel is shown as a gray fringe along the velocity axis. S/N in units of confidence level $\sigma$ are also plotted
across the radial velocity axis. Systemics uncertainty from the antenna noise, at a constant level among a single observing session, is shown in 
a box as $3\sigma$ error bars at the right border, referring to the 90 \% quantile of the line’s surface density value. The data were binned in frequency to 60 km s$^{-1}$ per channel. (c) APEX/SEPIA180 
spectrum centered at the redshifted CO(4-3) emission line frequencies from the lensed source, at redshift $z=1.4802$. The data were further binned in frequency, to 180 km s$^{-1}$ per channel, to try to increase the contrast} (same notation as for the central panel).
\label{fig:2}
\end{figure*}

\subsection{APEX Observations}\label{sec:apex}

We have observed this galaxy with APEX in order to confirm the origin of the strong submillimeter fluxes and the spectroscopic redshift of this emission.
For that we have carried out photometric and spectroscopic observations with APEX. We now describe these observations and the data reduction.

\subsubsection{APEX photometry} \label{sec:apex:pho}

\begin{figure*}
\gridline{
          \fig{fig_2a.pdf}{0.49\textwidth}{(a)}
          \fig{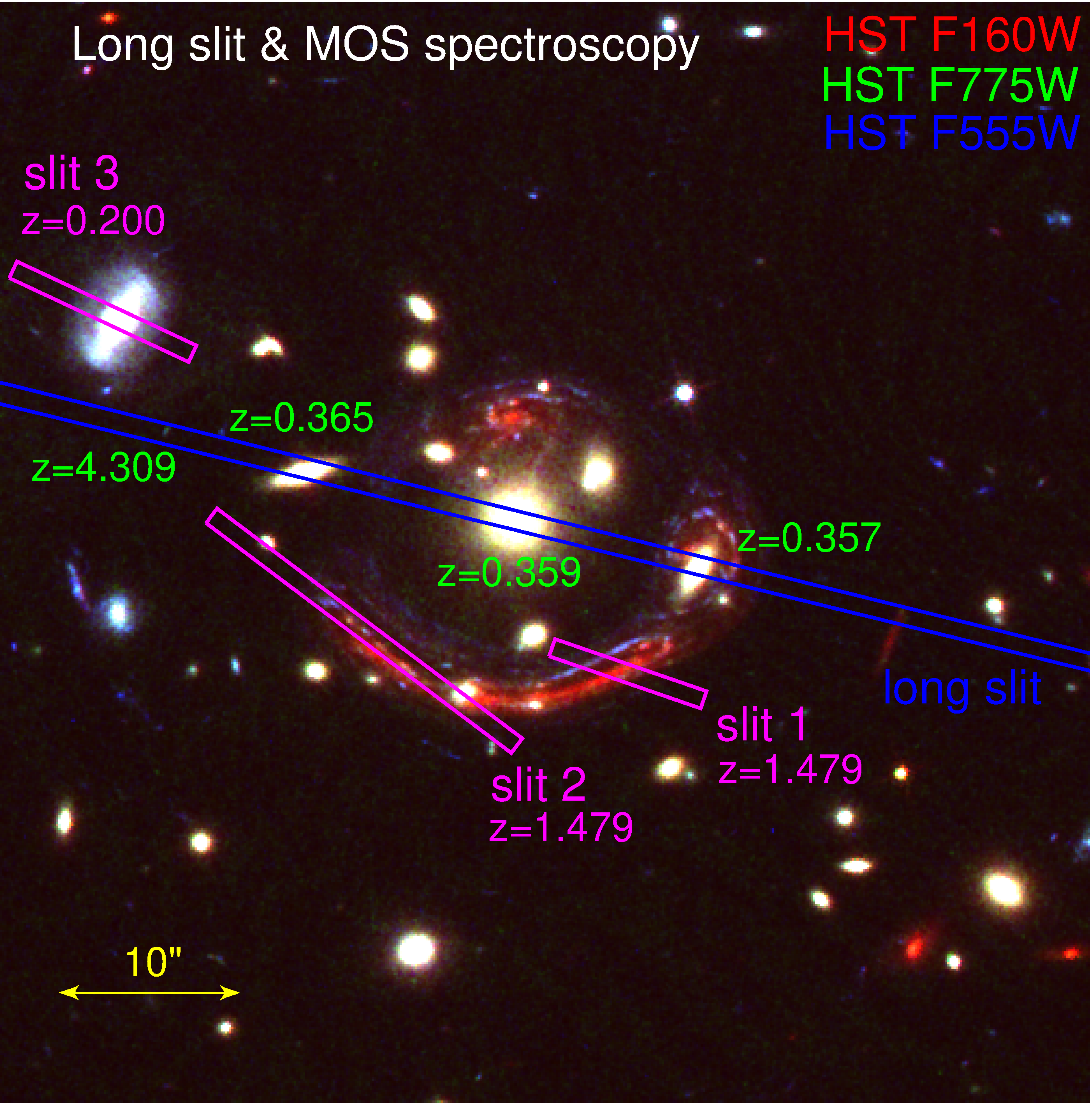}{0.492\textwidth}{(b)}
          }

 \caption{(a) $1\times 1$ arcmin$^2$ {\it HST/WFC3} RGB-image of GAL-CLUS-022058s with the filters, blue F555W, green F775W and red F160W. North is up; east is left. Yellow contours are from the APEX/LABOCA, contour levels are for $3\sigma$, $3.5\sigma$, $4\sigma$, $4.5\sigma$, $4.8\sigma$, and $4.9\sigma$. (b) RGB-image of GAL-CLUS-022058, which shows 3 slits from the VLT/FORS MOS (marked in magenta with their measured redshifts). The long slit (marked in blue) is from GEMINI GMOS observations. The obtained redshifts from the long slit are shown beside the galaxies in green.
}
\label{fig:hst}
\end{figure*}

Our observations were carried out between 18 and 19 April 2018, program ID 0101.B-0560(A) (PI Iglesias-Groth) using the bolometer LABOCA on APEX \citep{2009A&A...497..945S}.
We observed in mostly excellent weather conditions (PWV  0.5  mm and 0.9 mm). The data were reduced using the CRUSH \footnote{\url{https://www.sigmyne.com/crush/}} package, which is a free data reduction software for specific astronomical imaging arrays. It is especially designed for use with ground-based or air-borne (sub)millimeter wave cameras. Individual maps were co-added (noise-weighted) and the final map was beam smoothed, resulting in a spatial resolution with full width at half maximum (FWHM) of 19.5 arcsec. The total source observing time of the used data for this analysis is 8400 seconds and the average rms across the field is $8$ mJy beam$^{-1}$.
A submillimeter source is detected with coordinates consistent with those of {\it WISE} J022057.56-383311.4 which is partially responsible of the observed {\it Planck} submillimeter fluxes. We measure a APEX/LABOCA flux of $S_{870\mu m}= 54 \pm 8$ mJy. The differences with the expected {\it Planck} measurements (see Fig. \ref{fig:sed}) are due to the flux of the Sunyaev-Zeldovich ($SZ$) effect of a near galaxy cluster PSZ2 G249.80-68.11, at 4.5 arcmin north from our source at $z=0.228$ \citep{2002ApJS..140..239C}. In fact, the position of the source coincides with the arc structures of the strong lensed galaxy seen in the {\it HST} images as can be seen in Figure \ref{fig:hst}(a). In Figure \ref {fig:2}(a) we see the flux detection at S/N$\sim 6.6$. 

\subsubsection{APEX spectroscopy} \label{sec:apex:spec}

In order to obtain the redshift of the submillimeter emission we observed the submillimeter source with APEX program ID E-0105.20N6 (PI D\'iaz-S\'anchez). We first observed with APEX/SEPIA180 \citep{2018A&A...612A..23B} the CO(4-3) line at 186 GHz on 19 October 2020 with an exposure time of 47.2 minutes and PWV 2.1 mm. The CO(5-4) line was observed with APEX/nFLASH230\footnote{\url{http://www.apex-telescope.org/ns/nflash/}} at 232 GHz the days 23 October, 7 and 8 December 2020 with a total exposure time of 57.7 minutes and PWV 2.1-2.6 mm. The total time spent in this program was 5 hours including overheads.

Data reduction for the APEX spectra was performed with the software GILDAS\footnote{\url{https://www.iram.fr/IRAMFR/GILDAS}} with software package CLASS \citep{2005sf2a.conf..721P}. After collecting all daily observing scans, possible line signals are masked and polynomial baseline correction of order $n=5$ within an observed-frame frequency window of $\Delta\nu\approx1.5$ GHz centered on the signals was performed. Close to the CO(5-4) emission line, a spurious instrumental feature at $\nu_{obs}=232.1$ GHz is present in some of the scans. Thus contaminated channels were blanked. In order to suppress excess noise from inhomogeneous weather conditions across the daily observations, we ran a simple CLASS script \texttt{rmse\_selection\_function} that was specifically developed for blind spectroscopic line search by our group and is publicly available on GitHub\footnote{\url{https://github.com/NiSZR/rmse_selection_function}}. It works by optimally selecting individual scans according to the ranked baseline rms noise contribution in order to maximize the signal-to-noise content per coadded spectrum. Selected scans are averaged and the frequency channels are re-binned to source rest-frame 60 km s$^{-1}$ to compromise between sampling of the line profile and S/N-level for the full line detection. The flux conversion factor for APEX/SEPIA180 (nFLASH230) is 38.4$\pm$2.8 (35.0$\pm$3.0) Jy/K. The total uncertainty of the flux calibration is assumed to be less than 10 \%. Final flux calculations and visualization (see Figure \ref{fig:2}(b,c)) were performed in Python with modules NumPy \citep{2019zndo...3533894V} and astropy \citep{2018AJ....156..123A}. 

We have detected the CO(5-4) line emission with $\mathrm{S/N}=6$ obtaining a redshift of $z=1.4802$ with velocity-integrated line intensity of $I_{CO(5-4)}=21\pm 4$ Jy km s$^{-1}$ (Figure \ref{fig:2}(b)). The FWHM line width for CO(5-4) is $235 \pm 45$ km s$^{-1}$. We also have a tentative detection of CO(4-3) line with upper limit velocity-integrated line intensity of $I_{CO(4-3)} \approx 44$ Jy km s$^{-1}$ at $3\sigma$  (Figure \ref{fig:2}(c)). At this redshift, the flux for line CO(4-3) is shifted towards the edge of the 2 mm atmospheric window, causing strong, asymmetric noise contamination across the bandwidth.

\subsection{Archival HST Optical and NIR imaging} \label{sec:opt}

We have identified the arc structure  near to {\it WISE} J022057.56-383311.4 in archival {\it HST} images taken in 2015, proposal ID 13756 (PI Saurabh Jha), aimed at obtaining  {\it HST/WFC3} optical and infrared imaging of a large and nearly-complete Einstein ring. The optical images were taken using {\it HST/WFC3/UVIS} with filters F555W and F775W, with two exposures in each filter and a total exposure time of 1227 seconds and 1270 seconds respectively in 2015-05-15. The NIR image was taken with {\it HST/WFC3/IR} in filter F160W, with four exposures and a total exposure time of 2812 seconds in 2015-05-16. We have taken the raw data and processed them with the WFC3 pipeline \citep{Ge18} and IRAF package (Figure \ref{fig:hst}). There is not yet any scientific publication related this archival data.

\subsection{Archival VLT Spectroscopy} \label{sec:vlt}

\begin{figure}
   \centering
   \includegraphics[width=8.5cm]{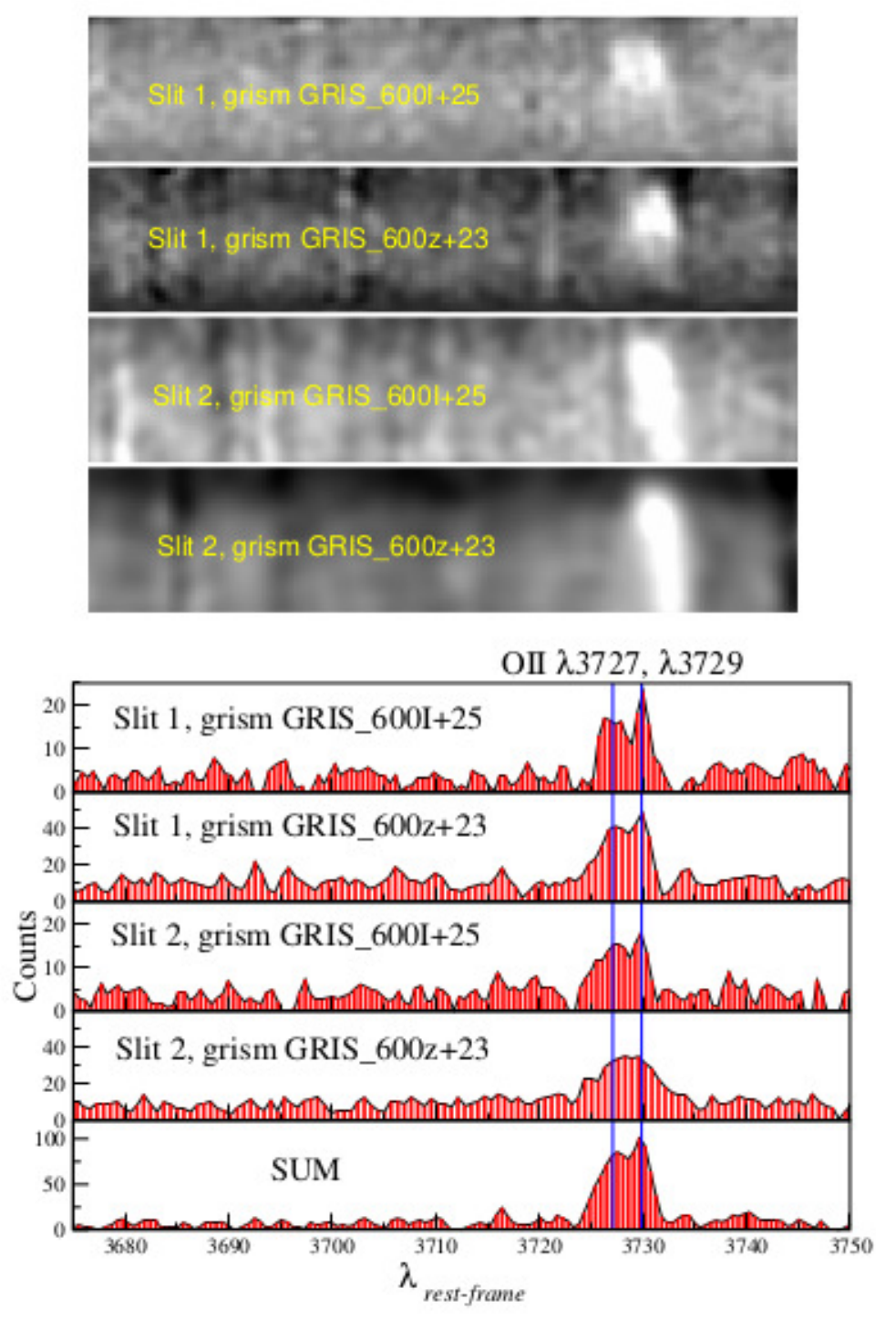}
   \caption{Top: The spectra for the two MOS slits on the arc structure, taken with two grisms (GRIS\_600I+25 \& GRIS\_600z+23), they are ordered as in the bottom panel. The doublet of OII $\lambda$3727, $\lambda$3729 \AA \ is detected in the two slits and for the two grisms, yielding a spectroscopic redshift $z=1.479$. Bottom: The spectra are shown as a function of the rest-frame wavelength, the sum of all spectra is also presented. Blue lines are for the double of OII.
    }
\label{fig:spt}
\end{figure} 

VLT/FORS2 observations were taken in 2008, ESO archive program ID 081.A-0693(A) (PI Graham Smith). For this cluster they designed a multi-slit mask containing 25 slits of 1.0 arcsec width to include as many of the candidate multiple systems as possible, with two tilted slits in the long arc structures (Figure \ref{fig:hst}(b)). Then they carried out 3600 seconds of exposure time with grism GRIS\_600I+25, and 2700 seconds of exposure using grism name GRIS\_600z+23.

We have processed and calibrated these archival data using IRAF package and our own specific code written in C++. 
Fig. \ref{fig:spt} shows the final results for the two slits and two grism in the spectral region where the doublet of OII $\lambda$3727, $\lambda$3729 \AA \ is detected. We have detected this doublet in the two slits on the arc structure and for the two grisms. From this doublet we obtain a spectroscopic redshift $z=1.479$ for the arc structures, which is consistent with the spectroscopic redshift obtained from APEX observations. 
For Slit 1 and 2 the doublet is detected in the whole region with significant flux, it is impossible distinguish if it comes from the bluest or the reddest clumps on the arc structures due to their separation (less than 1 arcsec). Considering the sum of the flux from both slits and grisms, we have $S/N > 10$ for the doublet of OII with a total exposure time of 1.75 hours, no more lines are detected in the observed range $\lambda_{obs}$=6700-10600 \AA \ (or $\lambda_{rest}$=2700-4270 \AA). We have measured the FWHM line width for the OII emission line (corrected from the instrumental profile) and we obtained $120 \pm 20$ km s$^{-1}$ for slit 1 and $145 \pm 20$ km s$^{-1}$ for slit 2. 

We have also measured redshifts for 19 more galaxies on mask slits. We show the redshift of these galaxies in Table \ref{tab:gal}. For galaxies with emission lines we obtained the redshift from lines H$\alpha$, NII $\lambda$6585\AA , and SII $\lambda$6718, $\lambda$6733 \AA , and for the galaxies without emission lines we used mainly the absorption lines Na $\lambda$5896 \AA , and Mg $\lambda$5177\AA ;  the H$\beta$ line was used when available. For a low redshift galaxy at $z=0.00009$, it was obtained from the triplet of CaII $\lambda$8498, $\lambda$8542, $\lambda$8662 \AA , absorption lines. A redshift of a galaxy of the cluster is found in the literature \citep{2009A&A...499..357G}.

\begin{figure*}
\gridline{\fig{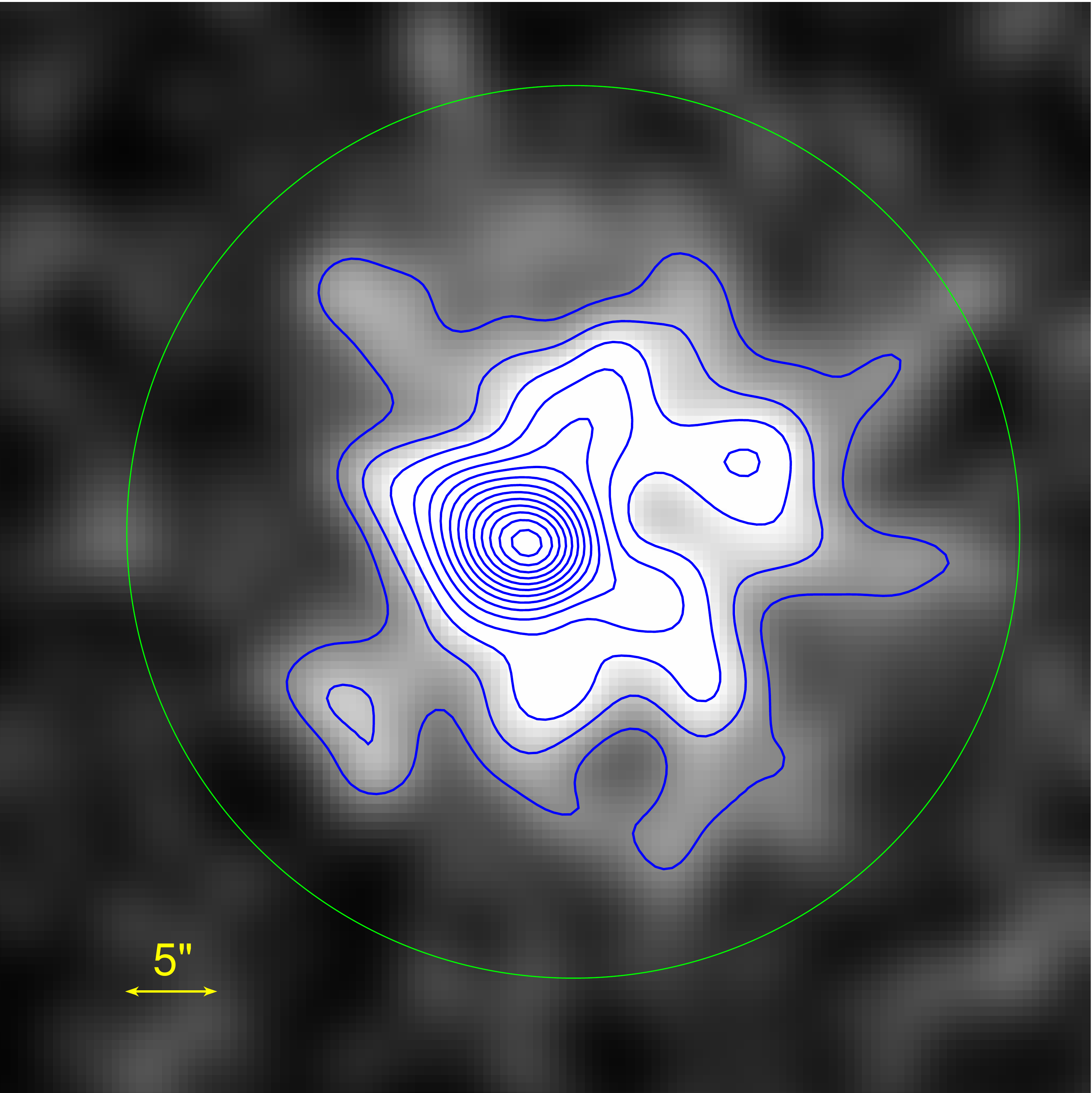}{0.48\textwidth}{(a)}
          \fig{fig_4b.pdf}{0.475\textwidth}{(b)}    
          }
   \caption{(a) $1\times 1$ arcmin$^2$ X-ray image smoothed with a 2 arcsec FWHM Gaussian. North is up; east is left. The first line of the contours is at 5$\sigma$ increasing lineally to 42$\sigma$, 14 levels with 2.6$\sigma$ separation among them. The radius of the circle is 25 arcsec and it is used to made aperture photometry. (b) $1\times 1$ arcmin$^2$ {\it HST/WFC3} RGB-image of GAL-CLUS-022058s, the X-ray contours are superimposed. North is up; east is left.
 }
\label{fig:rxg}
\end{figure*}

\begin{deluxetable}{cccccc}
\tablecaption{Spectroscopic redshift of GAL-CLUS-022058s \label{tab:gal}}
\tablecolumns{6}
\tablewidth{0pt}
\tablehead{
\colhead{RA} &
\colhead{Dec} &
\colhead{$z_{spec}$\tablenotemark{a}} &
\colhead{Class\tablenotemark{b}} &
\colhead{Telescope} &
\colhead{Type\tablenotemark{c}}
}
\startdata
02:20:57.27 & -38:33:12.2 & 1.47935 & SMG & VLT & E \\
02:20:58.36 & -38:33:10.7 & 1.47998 & SMG & VLT & E \\
02:20:59.50 & -38:32:52.0 & 0.20030 & G   & VLT & A \\
02:20:59.44 & -38:32:27.4 & 0.36418 & CG  & VLT & A \\
02:20:58.51 & -38:31:40.2 & 0.12377 & G   & VLT & E  \\
02:21:02.34 & -38:32:44.0 & 0.00009 & G   & VLT & A \\
02:20:59.96 & -38:31:17.7 & 0.23106 & G   & VLT & E \\
02:21:00.70 & -38:31:04.6 & 0.41227 & G   & VLT & A \\
02:20:59.87 & -38:30:13.1 & 0.22670 & G   & VLT & E \\
02:21:07.26 & -38:32:55.3 & 0.17080 & G   & VLT & E \\
02:21:02.93 & -38:30:33.8 & 0.16500 & G   & VLT & A \\
02:21:10.23 & -38:33:00.3 & 0.35784 & CG  & VLT & E \\
02:21:06.18 & -38:30:56.6 & 0.44211 & G   & VLT & A  \\
02:21:08.99 & -38:31:31.5 & 0.36089 & CG  & VLT & A  \\
02:21:09.05 & -38:30:59.5 & 0.23022 & G   & VLT & A \\
02:21:11.46 & -38:31:22.0 & 0.17824 & G   & VLT & E \\
02:21:12.54 & -38:31:27.8 & 0.23761 & G   & VLT & E \\
02:20:49.00 & -38:31:04.3 & 0.35851 & CG  & VLT & A  \\
02:20:52.37 & -38:33:09.6 & 0.36473 & CG  & VLT & A \\
02:20:54.46 & -38:35:23.8 & 0.36066 & CG  & VLT & A  \\
02:20:49.60 & -38:34:33.7 & 0.35551 & CG  & VLT & A  \\
02:20:56.88 & -38:33:06.0 & 0.35732 & CG  & GEMINI-S & A \\
02:20:57.73 & -38:33:03.1 & 0.35967 & BCG & GEMINI-S & E  \\
02:20:58.71 & -38:33:00.7 & 0.36574 & CG  & GEMINI-S & A  \\
02:20:59.36 & -38:32:58.4 & 4.30949 & G   & GEMINI-S & E  \\
02:20:53.79 & -38:27:38.1 & 0.36057 & CG  & Literature & E \\
\enddata
\tablecomments{$^a$Spectroscopic redshift. $^b$SMG=Submillimeter Galaxy, CG= Cluster Galaxy, BCG=Brightest Cluster Galaxy, G=Galaxy. $^c$E=Emission, A=Absorption.}
\end{deluxetable}

\subsection{Archival Gemini Spectroscopy} \label{sec:gemini}

Spectroscopic follow-up of the cluster of galaxies, using GMOS long slit with GEMINI-South in 2008, have been found in the GEMINI archive, archival program ID GS-2008A-Q-5 (PI Graham Smith). The observation were taken with GMOS-S, long slit of 1 arcsec width, grating B600+G5323, and 2400 seconds of exposure time.
The long slit used in the observations can be seen in Figure \ref{fig:hst}(b). We have processed these data and we have obtained the redshift of three galaxies of the cluster including the brightest cluster galaxy (BCG). The redshifts are shown in Table \ref{tab:gal}. Moreover, a galaxy with the redshift $z=4.309$ has also been detected.

For the BCG we obtain the redshift from the emission lines [OII] $\lambda$3726, $\lambda$3729 \AA , [OIII] $\lambda$5008 \AA , and H$\beta$, and the absorption lines K, H, and G. For the other two bright galaxies in the long slit we determine the redshift from the absorption lines K, H, and G. For the galaxy with the redshift $z=4.309$
the $Ly\alpha$ emission line is detected with $S/N \sim 15$ in 20 minutes of exposure time for two different set-ups of the instrument, and it is also detected in the {\it HST/F160W} band with AB-magnitude of 25.20, which is magnified a factor $\mu_{L}\sim 5$ (calculated with our lensing model in section \ref{sec:lens}).

\subsection{Archival X-Ray Observations} \label{sec:xray}

Archival {\it Chandra} observations were taken in 2008, archival program ID 9411 (PI Graham Smith), they observed a massive cluster of galaxies RXCJ0220.9-3829 at redshift $z=0.2287$ near of our galaxy cluster, at $\sim 4.5$ arcmin north. We can see in these observations X-ray emission coming from the galaxy cluster Fig. \ref{fig:rxg}. This X-ray emission have been mistaken with a point like source in \cite{2016ApJS..224...40W} with a flux of $F=5.79 \times 10^{-14}$ mW m$^{-2}$ but in the catalog given in \cite{2017ASPC..512..309E} there is a warning saying this is an extended source but they give the flux with an aperture of 2.4 arcsec which is $F=4.01 \times 10^{-14}$ mW m$^{-2}$.
We have taken the processed image in the energy band 0.5-7.0 keV, from the {\it Chandra} X-ray Center\footnote{\url{https://cda.harvard.edu/chaser/}}. In Fig. \ref{fig:rxg}(a) we show the smoothed image with a 2 arcsec FWHM Gaussian. We have made a 25 arcsec aperture radius photometry \citep{2007MNRAS.377..516D}, circle in green color in Fig. \ref{fig:rxg}(a), and obtained a flux of $F= (3.6 \pm 0.2) \times 10^{-13}$ mW m$^{-2}$. We show in Fig. \ref{fig:rxg}(b) the X-ray contours superimposed to the RGB-image.

\section{Analysis} \label{sec:ana}

\subsection{Galaxy Cluster} \label{sec:CG}

The lensed galaxy is gravitationally magnified by a massive foreground galaxy cluster at redshift $z=0.36$. 
We have obtained the spectroscopic redshift of 11 galaxies of the cluster. We have estimated the velocity dispersion of the cluster along the line-of-sight from the 11 galaxy members of the cluster with available spectroscopic redshifts (Table \ref{tab:gal}), which is $\sigma_{1\rm D}=684$ km s$^{-1}$. In order to estimate the size of the cluster we calculate the radius $R_{200}$ , which approximates the virial radius, from $\sigma_{1\rm D}$ \citep{2007MNRAS.377..516D} and we obtain
$R_{200}=1.39$ Mpc. We use the relation given in reference \cite{2013ApJ...772...47S} between the mass $M_{\rm cl}$ and $\sigma_{1\rm D}$ to calculate the cluster mass, which gives $M_{\rm cl}= (3.3\pm 0.5) \times 10^{14} M_\sun$. Assuming that the X-ray emission comes from the cluster of galaxies its X-ray luminosity is $L_X= (3.5 \pm 0.2) \times 10^{43}$ erg s$^{-1}$, which is compatible with its $M_{200}$ \citep{2011A&A...535A.105E,2020MNRAS.494.2736C}. We have measured the FWHM line width for two emission lines of the BCG, for the emission line [OII] $\lambda\lambda 3726,3728$ we obtain $373.2 \pm 20$ km s$^{-1}$, for the  emission line [OIII] $\lambda 5008$ we obtain $379.5 \pm 20$ km s$^{-1}$ and for H$\beta$ $396.1\pm 30$ km s$^{-1}$. These are narrow emission lines and we discard Type I AGN for the BCG.

\begin{figure*}
\gridline{
          \fig{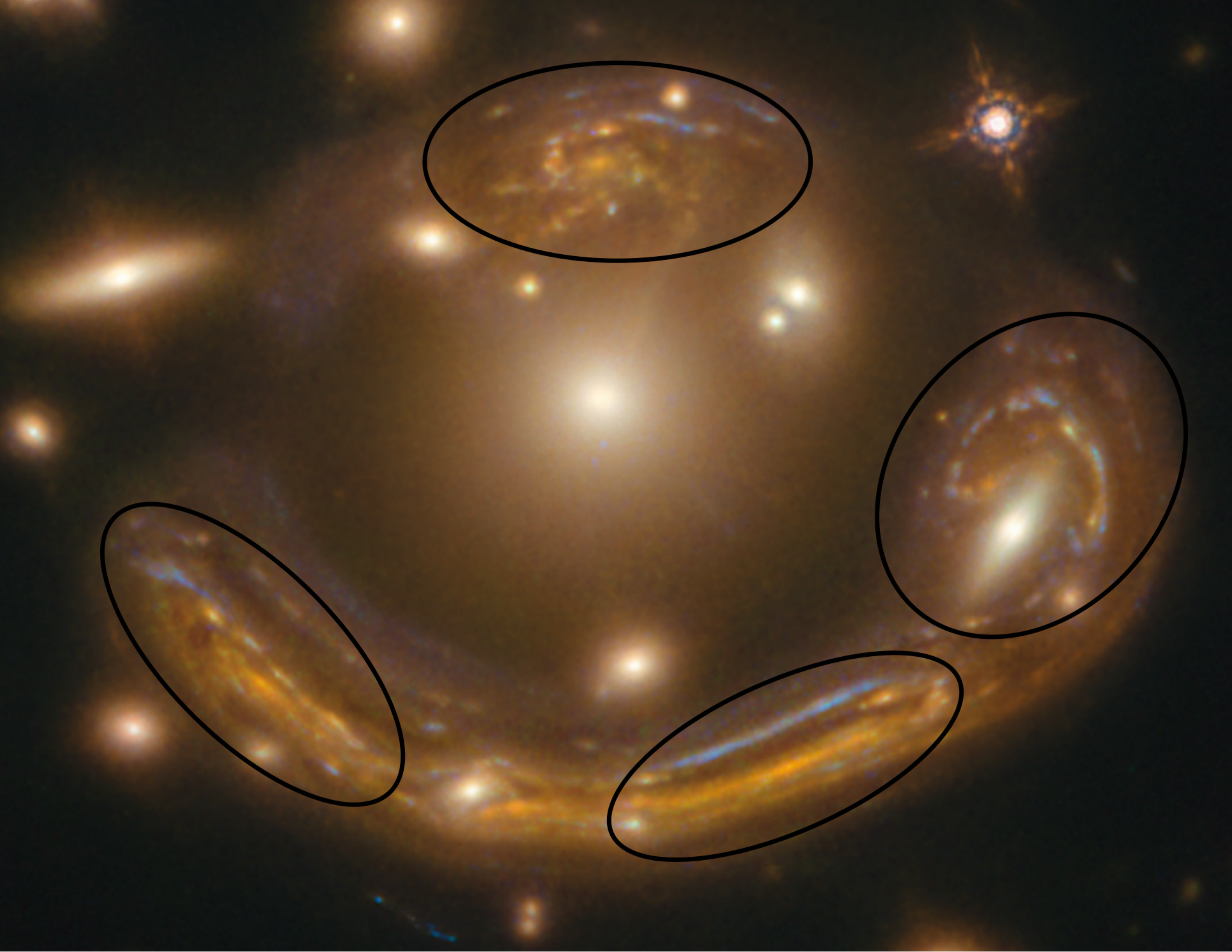}{0.525\textwidth}{(a)}
          \fig{fig_5b.pdf}{0.44\textwidth}{(b)}
          }

 \caption{
(a) $28\times 22$ arcsec$^2$ RGB-image of GAL-CLUS-022058s with the {\it HST/WFC3} filters. North is up; east is left. We mark the four images of the lensed galaxy we have used to calculate the lensing model (identifying clumps in these images). This image is taken from the image release as picture of the week by ESA/Hubble, where the authors named this object GAL-CLUS-022058s, also nicknamed ``Molten Ring"\textsuperscript{\color{blue} a }. (b) $ 30\times 30$ arsec$^2$ RGB-image of GAL-CLUS-022058s. The clumps used to model the gravitational lens in the four images of the galaxy are marked in yellow and there are four families ($a$, $b$, $c$, and $d$). Critical lines obtained from the lensing model for $z=1.48$ are drawn in blue.
}
\small\textsuperscript{\color{blue} a} \url{https://www.spacetelescope.org/images/potw2050a/}
\label{fig:lens}
\end{figure*}

\newpage 

\subsection{Lens modeling} \label{sec:lens}

We have used the public software Lenstool\footnote{\url{https://projets.lam.fr/projects/lenstool/wiki}} \citep{1993A&A...273..367K,2007NJPh....9..447J} to derive the lensing model. This software assumes a parametric model for the distribution of dark matter and performs a mass reconstruction of the foreground cluster. This model is constrained using the location of the multiple images identified in the cluster. We used a simple model with a single cluster-scale mass component, as well as individual galaxy-scale mass components centered on each cluster member. For each component, we used a dual pseudo-isothermal elliptical mass distribution (dPIE, also known as a truncated PIEMD \citep{2005MNRAS.356..309L}). Using SExtractor \citep{1996A&AS..117..393B} and visual inspection of the {\it HST/WFPC3} images, we have chosen multiple clumps with similar colors on the arc structure as family members of multiply lensed background components of the lensed galaxy at $z=1.479$. In selecting these image families we are supported by a preliminary lens model, which is based on the primary arcs \citep{2020ApJS..247...12S}. The initial model is successful in predicting the approximate locations of possible counter-images, in our case it has also predicted an elliptical mass distribution for the central halo. Using the theoretical results for this halo mass distribution we have identified four main images of the elliptical lensed galaxy. The lensing configuration is near to the so called ``fold arc'' \citep{2011A&ARv..19...47K} (see Fig. \ref{fig:lens}(a)). The western image is also distorted by the halo of a galaxy of the cluster.

Four families of four clumps were selected assuming redshift $z=1.479$ for all of them (see Fig. \ref{fig:lens}(b) and Table \ref{tab:fam}). 
Family $a$ is in the center of the elliptical galaxy and it is the reddest clump in the galaxy. Family $b$ is near to the central part and an arm of the galaxy and it has green-yellow color. And finally families $c$ and $d$ are on an arm of the galaxy and their colors are the bluest ones. For the eastern image of the galaxy we have measured the spectroscopic redshift with slit 2 (see section \ref{sec:vlt}), as the flux coming from the doublet of OII is in the whole region of the slit where the clumps are present, we have the spectroscopic redshift for the families of this image ($a.2$, $b.2$, $c.2$, and $d.2$). For the low central image of the lensed galaxy we have measured the spectroscopic redshift with slit 1 (see section \ref{sec:vlt}) and we have only flux coming from the doublet of OII for the central region of the galaxy ($a.3$) and for a blue arm ($c.3$).

We checked the model with the reconstruction of the arclets, and found good agreement between the positions for image clumps and the reconstruction for each component of the families (the rms between the original and the resconstructed positions is 0.41 arcsec). The data are best fitted with a cluster-scale potential of ellipticity $e = 0.172$, position angle PA =102.08$^{\rm o}$, normalization $\sigma_{\rm PIEMD} =796$ km s$^{-1}$, core radius $r_{\rm core} = 35.82$ kpc and cut radio $r_{\rm cut} = 1500$ kpc. The Einstein radius is $\theta_e \approx 10$ arcsec. 
The enclosed mass within $R_{200}=1.39$ Mpc is $M_{200}=3.8 \times 10^{14}M_\sun$, in concordance with the $M_{200}$ obtained in section \ref{sec:CG}. We can see the critical lines for $z=1.48$ in Fig. \ref{fig:lens}(b), the magnification for each clump is given in Table \ref{tab:fam}, obtained from the output of Lenstool software. The mean magnification in the center of the galaxy (family $a$) is $\mu_{\rm L}=18 \pm 4$ calculated from the four images of family $a$.

\begin{deluxetable}{ccccc}
\tablecaption{List of Lensing Constrains for the Lensing Model. \label{tab:fam}}
\tablecolumns{6}
\tablewidth{0pt}
\tablehead{
\colhead{ID \tablenotemark{a}} &
\colhead{RA} &
\colhead{Dec} &
\colhead{$\mu_{\rm L}$\tablenotemark{b}} &
\colhead{$z_{spec}$\tablenotemark{c}}
}
\startdata
a.1 & 35.2404838 & -38.5494009 & $16 \pm 4 $ &  \\
a.2 & 35.2432099 & -38.5529587 & $15 \pm 3$ &  1.47998 \\
a.3 & 35.2387213 & -38.5534089 & $25 \pm 6$ &  1.47935 \\
a.4 & 35.2365832 & -38.5512788 & $15 \pm 3$ & \\  
\hline
b.1 & 35.2404524 & -38.5491923 & $16 \pm 4$ & \\
b.2 & 35.2436567 & -38.5525034 & $12 \pm 2$ & 1.47998 \\
b.3 & 35.2399095 & -38.5535043 & $17 \pm 4$ & \\
b.4 & 35.2364258 & -38.5510711 & $12 \pm 2$ & \\
\hline
c.1 & 35.2400105 & -38.5490930 & $17 \pm 4$ & \\
c.2 & 35.2439491 & -38.5522620 & $10 \pm 2$ &  1.47998 \\
c.3 & 35.2393579 & -38.5532990 & $17 \pm 4$ &  1.47935 \\
c.4 & 35.2366492 & -38.5508668 & $14 \pm 3$ & \\
\hline
d.1 & 35.2397692 & -38.5491489 & $18 \pm 4$ & \\
d.2 & 35.2441465 & -38.5522513 & $9  \pm 1$ &  1.47998 \\
d.3 & 35.2384417 & -38.5529615 & $24 \pm 6$ & \\
d.4 & 35.2368640 & -38.5509108 & $18 \pm 3$ & \\
\enddata
\tablecomments{ $^a$The letter refers to the name of the group of images coming from the same region in the source galaxy. The number refers to the number of the image seen in the image plane. $^b$Magnification. $^c$Spectroscopic redshift.}
\end{deluxetable}

\subsection{SED} \label{sec:sed}

\begin{figure}
   \centering
   \includegraphics[width=8.5cm]{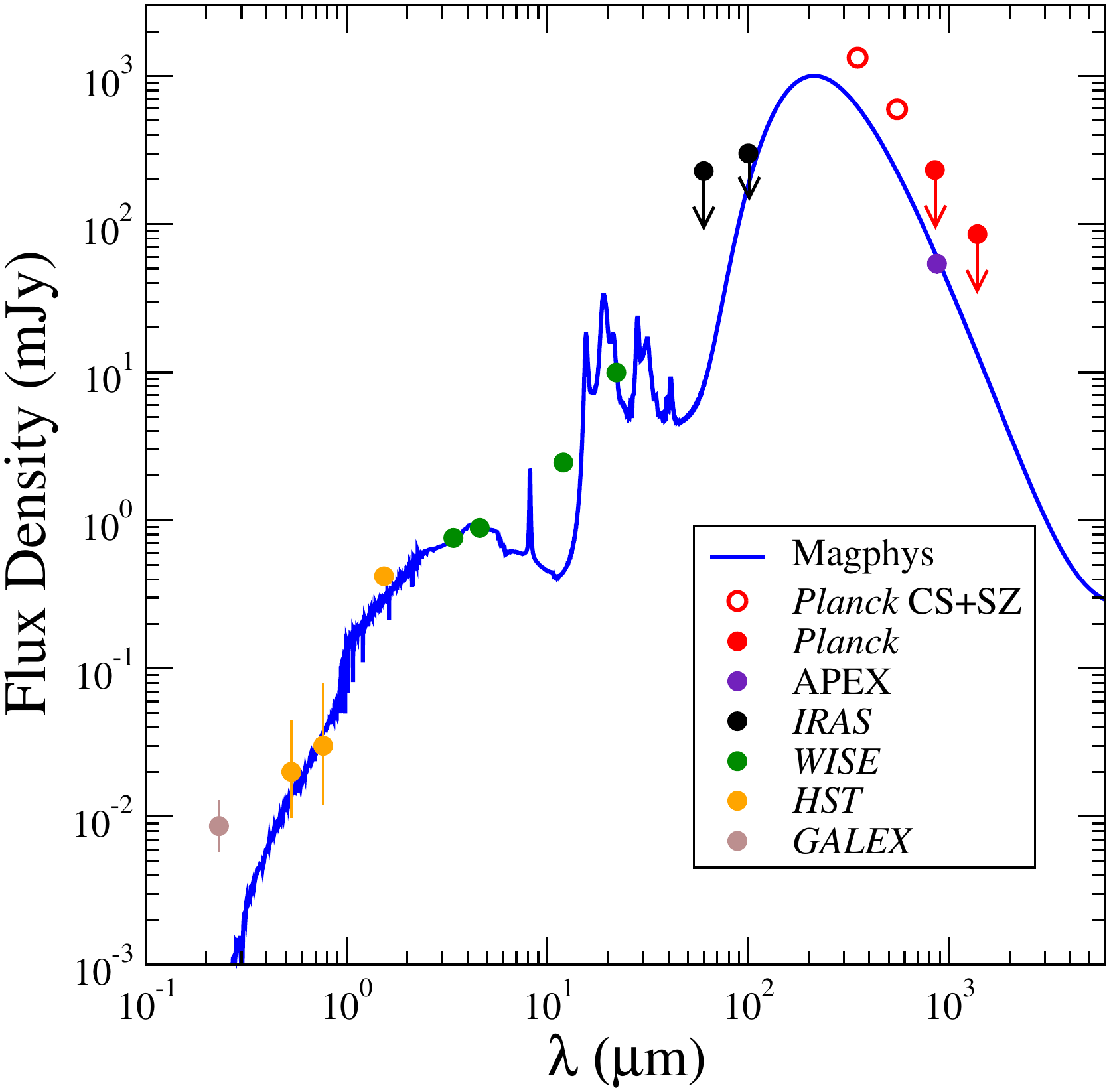}
   \caption{Multi-wavelength spectral energy distribution of GAL-CLUS-022058s. We have used the MAGPHYS code to fit the SED \citep{2008MNRAS.388.1595D}. The difference between the {\it Planck} measurements and the SED fit are due to the flux of the $SZ$ effect coming from a nearby galaxy cluster, PSZ2 G249.80-68.11, at 4.5 arcmin north from our source, and with $z=0.228$ \citep{2002ApJS..140..239C}. The fluxes from the compact source (CS) plus the ones from the SZ source are marked with red circles. 
    }
\label{fig:sed}
\end{figure}

We plot the SED (Fig. \ref{fig:sed}) and list the photometry in Table \ref{tab:sed}. We have also found data for our SMG in the {\it GALEX} NUV band \citep{2011Ap&SS.335..161B}. We have calculated the upper limit for 60 and 100 $\mu$m from {\it IRAS} Sky Survey Atlas \footnote{\url{http://irsa.ipac.caltech.edu/applications/IRAS/ISSA/}}. 
We have performed the photometry for {\it WISE} J022057.56-383311.4 using the {\it WISE} channels 3 and 4 from the {\it WISE} images. As there is no emission from any galaxy of the cluster for these channels, we have taken the whole measured flux in the region of the arc structures. For channel 4 there is only 0.4 magnitudes of difference with respect to the catalog magnitude and for channel 3 the difference is 1.2 magnitudes. For the other 2 channels we have taken the magnitudes given in the catalog for this source on the arc structure, in these channels another source on the BCG is detected and probably the flux for these bands is underestimated due to resolution and deblending process. For the {\it HST} images we have performed photometric measurements and taken the detected flux with S/N $> 3$ on the arc structures.

This SED is consistent with a source at redshift $z=1.479$ and brighter than other {\it Planck/Herschel} SMGs found so far \citep{2019ApJ...871...51F}. There is a cluster of galaxies with $SZ$ emission, PSZ2 G249.80-68.11, at 4.5 arcmin north from our source with $z=0.228$ \citep{2002ApJS..140..239C}. The flux detected by {\it Planck} is also coming from this source. We use the MAGPHYS code that allows us to fit simultaneously the ultraviolet-to-FIR SED, so we can constrain physical parameters \citep{2008MNRAS.388.1595D}. We assume that the submillimeter flux is coming from a region near to the center
of the lensed galaxy and we take the mean amplification for family $a$ as the submillimeter amplification, which is $\mu_{\rm L} =18$.
The fit is shown in Figure \ref{fig:sed}. From this fit we have the following physical parameters corrected for lensing amplification and image multiplicity, stellar mass log$(M_\ast/M_\sun) = 10.7 \pm 0.1$, specific star formation rate (sSFR) log$(sSFR/Gyr^{-1})=0.15 \pm 0.03$,
average V-band dust attenuation $A_V = 2.30 \pm 0.01$, luminosity-averaged dust temperature $T_{\rm dust}/K = 36 \pm 2$, mass-weighted age log$(age_M\ yr^{-1})= 8.84 \pm 0.02$, dust luminosity log$(L_{\rm dust}/L_\sun) = 12.0 \pm 0.1$, dust mass log$(M_{\rm dust}/M_\sun) = 8.12 \pm 0.09$, and star formation rate (SFR) log$(SFR/M_\sun\ yr^{-1}) = 1.83 \pm 0.02$.
We also calculate the SFR from the dust luminosity and the expression given in \cite{1998ApJ...498..541K} $SFR \sim 172$ $M_\sun\ yr^{-1}$, which assumes a Salpeter IMF, assuming a Chabrier IMF would possibly be a factor of $\sim$ 1.8 lower \citep{2014PhR...541...45C}.

The sSFR agrees with the so-called main sequence of star-forming galaxies with a stellar mass of $10^{11}$ M$_\sun$ at this redshift \citep{2021ApJ...908..192S}, and if we compare the sSFR with the dust-to-stellar mass ratio (also referred to as the specific dust mass), this correlation also matches with previous high redshift SMGs \citep{2021ApJ...908..192S}. The SFR dependence with the stellar mass is consistent with the main sequence behavior \citep{2021ApJ...910L...7C,2014ApJ...795..104W,2017ApJ...840...78D}.

\begin{deluxetable}{lcc}[b!]
\tablecaption{Ultraviolet to Submillimeter photometry of GAL-CLUS-022058s \label{tab:sed}}
\tablecolumns{3}
\tablewidth{0pt}
\tablehead{
\colhead{Wavelength($\mu$m) } &
\colhead{Flux (mJy)\tablenotemark{a,b}} &
\colhead{Observatory/Instrument}
}
\startdata
0.23157 & 0.009 $\pm 0.003$ & $GALEX/NUV$ \\ 
0.5308 &  0.02 $\pm 0.01$ & $HST/WFC3$ \\
0.7647 & 0.03 $\pm 0.02$ & $HST/WFC3$  \\
1.5369 &   0.42 $\pm 0.02$ & $HST/WFC3$ \\
3.4 & 0.76 $\pm 0.02$ & $WISE$ \\
4.6 &  0.89 $\pm 0.02$ & $WISE$ \\
12 &  2.45 $\pm 0.08$ & $WISE$ \\
22 & 9.9 $\pm 0.7$ & $WISE$ \\
60 & $<$ 200$^{c}$ & $IRAS$ \\
100 & $<$ 300$^{c}$ & $IRAS$ \\
350 & 1400 $\pm 600$$^d$ & $Planck$ \\
550 & 800 $\pm 300$$^d$ & $Planck$ \\
850 & $<$ 200$^{c}$ & $Planck$ \\
870 & 54 $\pm 8$ & APEX/LABOCA \\
1400 & $<$ 50$^{c}$ & $Planck$ \\
\enddata
\tablecomments{
$^a$Uncorrected for lensing amplification. $^b$Total arc flux. $^c$Upper flux limit. $^d$ $SZ$ + Compact Source fluxes.
}
\end{deluxetable}

\subsection{CO lines} \label{sec:CO}

Figure \ref{fig:2}(b,c) shows the CO line detections at the position of our SMG. We have identified two lines CO(5-4) and CO(4-3). The binned $\Delta V= 60$ km s$^{-1}$ CO(5-4) line is detected with flux density $S_{CO(5-4)} = 40 \pm 6$ mJy, velocity-integrated line intensity $I_{CO(5-4)} = 21 \pm 4$  Jy km s$^{-1}$, FWHM line width $235 \pm 45$ km s$^{-1}$ and redshift $z=1.4802$. The binned $\Delta V= 180$ km s$^{-1}$ CO(4-3) tentative detection line would have $S_{CO(4-3)} \approx 93$ mJy and $I_{CO(4-3)} \approx 44$  Jy km s$^{-1}$, at the redshift $z= 1.4802$. This last line is located close to a strong atmospheric H$_2$O line at 183 GHz. Accounting for the low $S/N \sim 3$ of this line measurement, the inferred line intensity should be considered as an upper-limit.
 
For the CO(5-4) line we derive a luminosity $\mu_{\rm L}L'_{CO(5-4)} = 1.0\pm 0.2 \times 10^{11}$ K km s$^{-1}$ pc$^2$, which is uncorrected from lensing amplification and image multiplicity. If we compare this luminosity with the also uncorrected IR luminosity, log$(\mu_{\rm L}L_{\rm IR}/L_\sun) = 13.85$, it agrees with the expectation for high redshift SMGs \citep{2013ARA&A..51..105C}. The luminosity of this line is high and it can be compared with the ones listed in \cite{2021ApJ...908...95H}, 5 from the 24 SMGs listed in this reference have a lower CO(5-4) luminosity,  $\mu_{\rm L}L'_{CO(5-4)}$.

\cite{2013MNRAS.429.3047B} calculated average flux ratios for a sample of 40 non-lensed luminous SMGs at comparable redshift. Demanding similar subthermal conditions as in this reference and small differential lensing effects, the velocity-integrated line intensity for CO(1-0) line can be derived,
\begin{equation}
I_{CO(1-0)} \approx \frac{I_{CO(5-4)}}{25 \langle r_{2,1}\rangle \langle r_{3,2}\rangle \langle r_{4,3}\rangle \langle r_{5,4}\rangle } = 14.7 {\rm\ Jy\ km\ s^{-1}} .\label{ecu:co}
\end{equation}

We use the CO luminosity to molecular gas mass conversion factor $\alpha_{\rm CO}=0.8$ $M_\sun$ pc$^{-2}$ (K km s$^{-1}$)$^{-1}$, a commonly used value for merger-induced star
formation \citep{2005ARA&A..43..677S}. Including a factor of 1.36 to account for helium, and correcting for lensing amplification and image multiplicity, we derive M$_{\rm mol}$ $\approx 2.6 \times 10^{10} M_\sun$. This value agrees with the main sequence relation of the star-forming galaxies for SFR-M$_{\rm mol}$ \citep{2014ApJ...793...19S}. For galaxies on the main sequence, the gas mass conversion factor $\alpha_{\rm CO}$ is larger than the one used here, up to $\sim 5$ larger \citep{2012A&A...548A..22M}. For the 24 {\it Planck} lensed SMGs listed in \cite{2021ApJ...908...95H} with similar gas mass, they found great dispersion, $\alpha_{\rm CO}=0.8-4$. On the other hand, the conversion factor given in equation \ref{ecu:co} also has a great uncertainty. Nevertheless, the $M_{\rm mol}$ value obtained here agrees with the one calculated from the dust continuum in the next section, but it could be between M$_{\rm mol}$ $\approx (2.6-13) \times 10^{10} M_\sun$.

\section{Discussion}\label{sec:dis}

Cross-matching between the {\it AllWISE} and {\it Planck} full-sky compact source catalogs we have identified a bright high redshift lensed SMG. It is lensed by a galaxy cluster at $z=0.36$ with a mean amplification factor in the center of the galaxy of $\mu_{\rm L} \approx 18 $, four images of the lensed galaxy have been identified. The redshifts derived from the submillimeter spectra and the optical one agree with each other very well, indicating that the submillimeter flux is coming from the arc structure at $z \approx 1.48$.
The uncorrected IR luminosity is log$(\mu_{\rm L}L_{\rm IR}/L_\sun) = 13.85$ and it is among the brightest high redshift lensed galaxies found with {\it Planck} \citep{2021ApJ...908...95H}. The intrinsic luminosity is $L_{\rm IR} \approx 10^{12}$ $L_\sun$, which gives a SFR of $\sim 70-170$ $M_\sun\ yr^{-1}$. We assume that the submillimeter flux is coming from a region near to the center of the lensed galaxy and we take the mean amplification for family $a$ as the submillimeter amplification. The amplification depends on the position of the flux in the image plane, see Table \ref{tab:fam}, if it was near to the critical lines then the amplification would be much larger. We correct for lensing magnification and image multiplicity, and with $T_{\rm dust}= 36$ K and the observed 850 $\mu m$ dust continuum, we use Equation (16) in \cite{2016ApJ...820...83S} to derive M$_{\rm mol} =1.7 \pm 0.9 \times 10^{10} M_\sun$. This value agrees with the main sequence relation of the star-forming galaxies for SFR-M$_{\rm mol}$ \citep{2014ApJ...793...19S} and with the value obtained from the CO(5-4) line calculated in section \ref{sec:CO}.
We have derived physical properties from the SED using MAGPHYS. Comparing the stellar mass with the star formation rate, we find that this galaxy lies on the main sequence of the star-forming galaxies at $z\sim 1.5$ \citep{2021ApJ...910L...7C,2014ApJ...795..104W,2017ApJ...840...78D}. The sSFR also agrees with the main sequence of star-forming galaxies at this redshift \citep{2021ApJ...908..192S,2020MNRAS.494.3828D}. The comparison between the sSFR and the dust-to-stellar mass ratio also matches with previous high redshift SMGs \citep{2021ApJ...908..192S}. From the luminosity for the CO(5-4) line, $\mu_{\rm L}L'_{CO(5-4)}$, and the uncorrected IR luminosity $\mu_{\rm L}L_{\rm IR}$, this galaxy compares well with the main sequence of star-forming galaxies \citep{2013ARA&A..51..105C}. The gas fraction is $f=M_{\rm mol}/(M_{\rm mol}+M_\ast)=0.34 \pm 0.07$, which agrees with the expectations at this redshift for SMGs \citep{2020MNRAS.494.3828D,2013MNRAS.429.3047B}.

We have measured different widths for the [OII] and C(5-4) lines, one is detected in the optical and the other one in the submillimeter regime, so it is possible that these emissions would come from different regions in the galaxy, i.e. the optical line would come from the spiral arms and the submillimeter line from the center of the galaxy near to an AGN. The submillimeter emission would come from the interaction of dense ISM around a central supermassive black hole, so it is wider than the optical one.

\section{Conclusion}\label{sec:concl}

We have found an ultra-bright SMG at $z_{spec}=1.4796$ cross-matching the {\it AllWISE} and {\it Planck} catalogs. Our APEX/LABOCA observations give the position of the submillimeter flux which coincides with the position of a multiple lensed galaxy, the Einstein ring GAL-CLUS-022058s, observed with {\it HST}. We have obtained the spectroscopic redshift from archival VLT/FORS observations and confirmed the redshift  of the submillimeter emission with our APEX/nFLASH230 observations. The lensed source appears to be gravitationally magnified by a massive foreground galaxy cluster lens at $z = 0.36$, we have identified four images of the lensed galaxy. We have used Lenstool to model the gravitational lensing which is near to a ``fold arc''  configuration for an elliptical mass distribution of the central halo; the mean magnification is $\mu_{\rm L} =18\pm 4$. We have determined an intrinsic rest-frame infrared luminosity of $L_{\rm IR} \approx 10^{12} L_\sun$ and a likely SFR of $\sim 70-170$ $M_\sun\ yr^{-1}$. The molecular gas mass is $M_{\rm mol} \sim 2.6 \times 10^{10} M_\sun$ and the gas fraction is $f = 0.34\pm 0.07$. We also obtain a stellar mass log$(M_\ast/M_\sun) = 10.7 \pm 0.1$ and a specific SFR log$(sSFR/Gyr^{-1})=0.15 \pm 0.03$. 
This galaxy lies on the so-called main sequence of the star-forming galaxies for several parameters. 
Finally we emphasize that a large amount of data used here are from archive. Our new observations have been carried out with APEX telescope.

\newpage 

\acknowledgments

This publication is based on data acquired with the Atacama Pathfinder Experiment (APEX) under programmes ID 105.20N6 and ID 0101.B-0560(A). APEX is a collaboration between the
Max-Planck-Institut fur Radioastronomie, the European Southern Observatory, and the Onsala Space Observatory. We thank Carlos De Breuck, and the APEX team for their support. Based on observations collected at the European Southern Observatory under ESO programme ID 081.A-0693(A). Based on observations made with the NASA/ESA Hubble Space Telescope, and obtained from the Hubble Legacy Archive, which is a collaboration between the Space Telescope Science Institute (STScI/NASA), the Space Telescope European Coordinating Facility (ST-ECF/ESA) and the Canadian Astronomy Data Centre (CADC/NRC/CSA). This publication makes use of data products from the Wide-field Infrared Survey Explorer. Based on observations obtained at the international Gemini Observatory program ID GS-2008A-Q-5.  The scientific results reported in this article are based in part on data obtained from the Chandra Data Archive. This work has been partially funded by projects ``Participation in the NISP instrument and preparation for the scientific exploitation of Euclid", PID2019-110614GB-C22/AEI/10.13039/501100011033,  \& PID2019-110614GB-C21 financed by the ``Agencia Estatal de Investigación'' (AEI-MCINN). H.D. acknowledges financial support from the Spanish Ministry of Science, Innovation and Universities (MICIU) under the 2014 Ramón y Cajal program RYC-2014-15686 and under the AYA2017-84061-P, co-financed by FEDER (European Regional, Development Funds), and in addition, from the Agencia Estatal de Investigación del Ministerio de Ciencia e Innovación (AEI-MCINN) under grant (La evolución de los cíumulos de galaxias desde el amanecer hasta el mediodía cósmico) with reference (PID2019-105776GB-I00/DOI:10.13039/501100011033). N.S. acknowledges a research stay at the Instituto de Astrofísica de Canarias, funded by ``Short-term grant abroad" (KWA) programme of the University of Vienna.



\newpage

\bibliography{apj_GALCLUS_resub_4_PR}{}
\bibliographystyle{aasjournal}

\end{document}